\documentclass[12pt]{nih2}

\usepackage[pdftex]{graphicx}
\usepackage{fancyhdr}
\begin{document}
\noindent
\begin{center}
\vspace*{-0.281in}
\LARGE
\textbf{Solute Transport Within Single Pores Post-Pulse} \\
\normalsize
\vspace{0.09in}
Julie V. Stern$^1$, Thiruvallur R. Gowrishankar$^1$, and James C. Weaver$^{1,*}$\\

\vspace{0.12in}
$^1$Harvard-MIT Division of Health Sciences and Technology, Institute for Medical Engineering and Science, Massachusetts Institute of Technology, Cambridge, MA, USA;
$^*$Corresponding author: jcw@mit.edu\\

\vspace{0.12in}
\end{center}

\vspace{0.23in}
\Large
\textbf{Abstract} 
\vspace{0.12in}
\normalsize

\textbf{We present results of studying the post-pulse transport properties of single pores.  Single pores exhibit active transport (drift), and not just passive transport (diffusion) at early post-pulse times.  In addition, we suggest experimental design methods obtained from model parameters that would enable experimentalists to find localized regions of a few pores.  
}

\vspace{0.23in}
\Large
\textbf{Introduction} 
\vspace{0.12in}
\normalsize

\large
\textbf{Why Study Single Pores?} 
\vspace{0.12in}
\normalsize

Single pores represent unique challenges to the study of electroporation in lipid bilayer membranes.  The main reason to study singles pores is to understand the physical underpinnings of transport and electrical behavior at the most basic level. Before investigating a cell with multitudes of pores sometimes numbering in the 10s of 1000s, it is always advantageous to study basic building blocks, 1 or a few pores.  With the lack of ability to observe pore structure, these 'dark pores' \cite{2017arXiv170807478W} are modeled functionally, the results showing how the physics plays out leaving clues to the underlying mechanisms.

This study concentrates on post-pulse behavior as that poses interest from the perspective of accessible time frames that experimentalists may reach.  In addition, the dynamics of diffusion and drift and which is dominant can best be addressed post-pulse. 
S\"{o}zer et. al. 2017 \cite{SozerLevineVernier_QuantitativeLimits_SmallMoleculeTransport_Electropermeome_MeasuringModeling-SingleNanosecondPerturbations_SciRep2017} reports that diffusion alone cannot be the only mechanism at play during post-pulse time frames.  

Another major area of interest in embarking on the study of single pores is that 1 pore may behave differently than a collective.  In addition, Stern et. al. 2017 \cite{2017arXiv170807613S} asserts that there is a distribution of pore lifetimes which suggests the possibility of multiple pore types.  Once we can study 1 pore, we can extend study to the different types of single pores and isolate varying characterstics.    

Studying single pores also pushes the cell model to extremes where naturally some limits may be reached.  In the same light, experimental detection in this regime also has its limits.  Hence, this study attempts to share some insights from modeling that
can be applied to understanding and reaching experimental detection limits.

\vspace{0.12in}
\Large
\textbf{Methods} \\
\normalsize

\large
\textbf{The Cell Model and Geometrical Aspects} 
\vspace{0.12in}
\normalsize

The model used is the circular cell model by Smith \cite{SmithKC_CellModelWithDynamicPoresAndElectrodiffusionofChargedSpecies_DoctorateThesisMIT_2011}. Figure 1 illustrates the regional domains that separate the inside and outside of the cell, and shows the full simulation box with its mesh as well as a zoom in of the membrane.  The membrane consists of 150 transmembrane node pairs, with a mesh node on each side of the membrane.  A typical cell size is radius 5 $\mu$m.  A typical membrane thickness is 4nm.  The underlying mathematical framework forms a physically based continuum model.  This model describes molecular transport and its flux properties.  The model uses a pore energy landscape which defines pore creation rates and their energy barriers.  The model can also determine local transmembrane voltage, pore distribution, conductance, hindrance, and the partitioning of solutes and ions into the pores. 

\begin{figure}
\begin{center}
\begin{tabular}{ccc}
\includegraphics[width=2.2in]{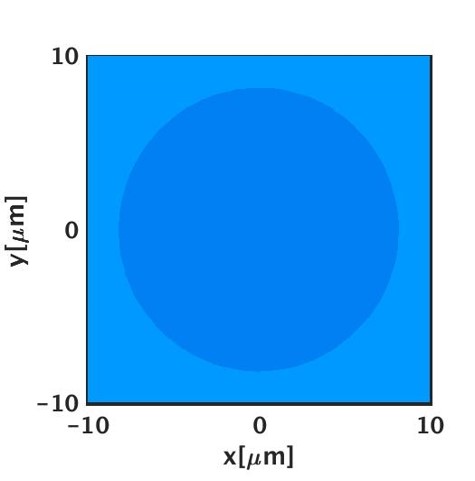} &
\includegraphics[width=2.2in]{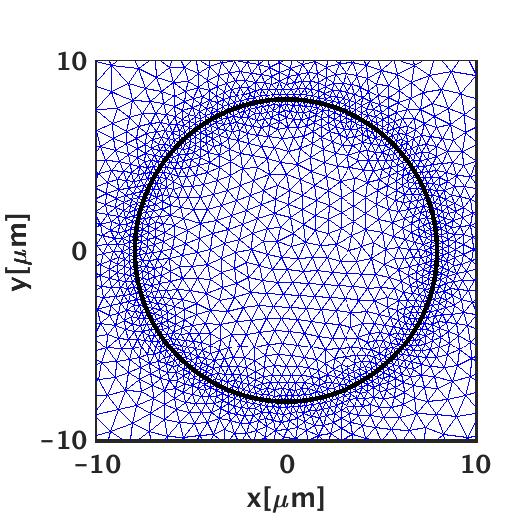} &
\includegraphics[width=2.2in]{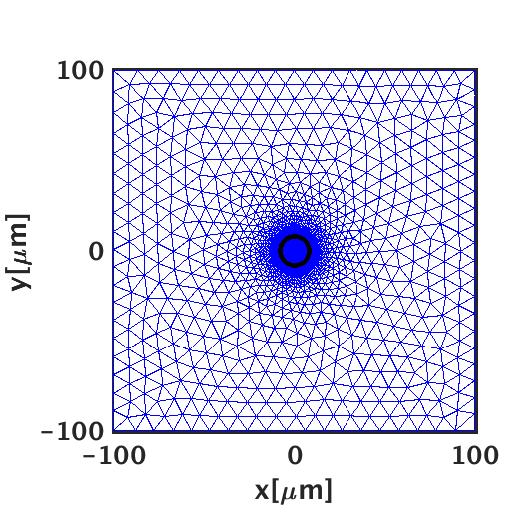} \\
 &  & \\
\end{tabular}
\end{center}
\caption{\textbf{Geometry and Model.}
Intracellular and extracellular (left) regions for solutes and ions.  Membrane and mesh (middle).  200 $\mu$m x 200 $\mu$m simulation box domain with mesh.  }
\end{figure}

\vspace{0.12in}
\large
\textbf{Modeling Aspects of a Pore} 
\vspace{0.12in}
\normalsize

In the Smith model \cite{SmithKC_CellModelWithDynamicPoresAndElectrodiffusionofChargedSpecies_DoctorateThesisMIT_2011}, a pore is modeled with a trapezoidal geometry and includes a cylindrical interior.  The
height of the pore corresponds to the membrane thickness.  A pore has a resistance based on three resistances in series, the interior resistance and two outer spreading resistances.  The spreading resistance functions like a 'focusing field'.  Ions follow the pathway of the electric fields in the extracellular bath.  Since the only pathway an ion has to cross the membrane is through the pore, it is as if the fields are 'focused' \cite{SmithKC_CellModelWithDynamicPoresAndElectrodiffusionofChargedSpecies_DoctorateThesisMIT_2011} at openings of the pore.
Figure 2 (left) illustrates this.

As the electric field is applied to the cell, a total membrane capacitance arises.  The cell
sits within the field.  The anodic pole of the cell is closest to the positive side of the charge separation, and the cathodic pole is closest to the negative side.  The pore is more likely to develop on the anodic pole side of the cell as depicted in the schematic in Figure 2 (right). 

In the model, pores can be fractional since they are modeled to be probabilistic. Therefore, one single pore can be represented by fractional pores which sum to one.

\begin{figure}
\begin{center}
\begin{tabular}{cc}
\includegraphics[width=3.2in]{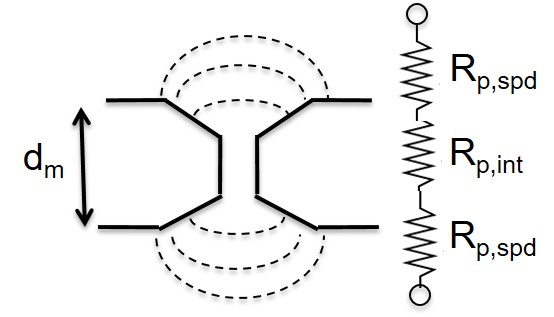} & 
\includegraphics[width=2.0in]{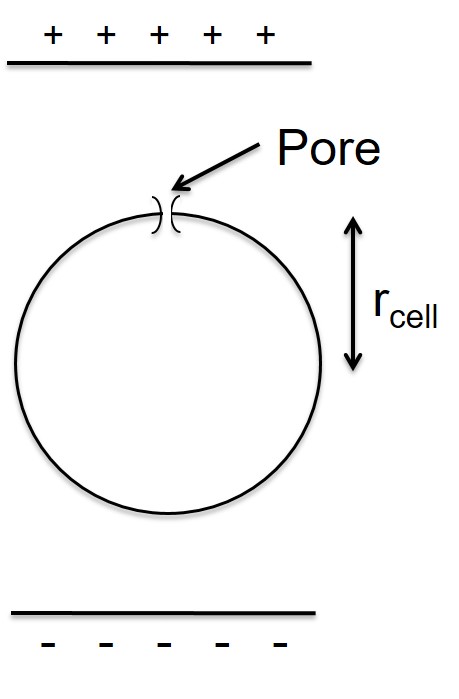} \\
$d_m$ membrane thickness & $C_m$ total membrane capacitance \\
\end{tabular}
\end{center}
\caption{\textbf{Pore Model Characteristics.}
Pore shape and resistances (left).  Pore positioning on cell within electric field (right).  Not drawn to scale.}
\end{figure}

\vspace{0.12in}
\large
\textbf{Iterative Methods and Sensitivity to Applied Electric Fields} 
\vspace{0.12in}
\normalsize

An iterative algorithm is applied which results in determining the electric field strength for the applied pulse of the trapezoidal waveform.  The algorithm starts off with a guess for applied electric field strength and sets upper and lower boundaries on field strength.  The number of pores generated from a pulse of this field strength is calculated and if it is greater than our allowable margin of error, we iterate to the next guess, while the upper and lower bounds are adjusted.  If the guess was too high, then that becomes the new upper bound, and if too low, the new lower bound.  The next guess is midway between the newly adjusted upper and lower bounds.  This way the electric field strength for the applied pulse can be found quickly, and within an acceptable margin of error.

\vspace{0.12in}
\large
\textbf{Model Parameters} 
\vspace{0.12in}
\normalsize

     The approach used in this study was to compare the single pore results with results from a control case containing a large number of pores.  The large pore case chosen was based on the parameters used in Kennedy et. al. 2008 \cite{KennedyEtAlBooske_QuantificationElectroporationKineticsPI_FinalVersion_BPJ2008}.  Both 0 mV and -50 mV resting potential initial conditions were run.  The large pore case had $\sim$5700 pores for an initial resting potential of -50 mV and $\sim$5600 pores for 0 mV initial resting potential.  In addition, results for 10 pores were generated to be an example of a 'few' pores.
      The model parameters used based on Kennedy et. al. 2008 \cite{KennedyEtAlBooske_QuantificationElectroporationKineticsPI_FinalVersion_BPJ2008} can be found in Table 1.  The solute used in this study is propidium ion (Pr++).

\begin{figure}
\begin{center}
\begin{tabular}{c|c}
\multicolumn{2}{c}{{\bf Table 1:  Model Parameters}}\\
\multicolumn{2}{c}{\ }\\
Parameter & Description\\ \hline
Fixed Membrane tension & 1e-05 N/m  \\
Pulse Duration & 40us  \\
Pulse rise and fall & 1us  \\
Pore Lifetime & 4s  \\
Extracellular Initial Pr++ & 30uM  \\
Intracellular Initial Pr++ & none  \\
Waveform type & Trapezoid \\
\end{tabular}
\end{center}
\end{figure}

	Table 2 contains the actual data obtained as a result of the simulation and iterative process of determining the electric field strengths corresponding to a single pore, 10 pores, and 5700 pores.  As shown, the margin of error on maximum pore number was less than 1.25\% for the Eapp of 2.05 kV/cm and less than .1\% for the 10 pore cases, and less than 1\% for the single pore cases.  Hereafter, the pore number will be referred to as its nearest whole number listed in the table.

\begin{figure}
\begin{center}
\begin{tabular}{c|c|c|c}
\multicolumn{4}{c}{{\bf Table 2:  Applied Electric Field Strength (Eapp) and Pore Number}}\\
\multicolumn{4}{c}{\ }\\
Um\_{0} & Eapp & N\_{max} & N\_{max\_orig} \\ \hline
-50 mV & 2.05 kV/cm & 5700 & 5731.2 \\
-50 mV & 0.9168 kV/cm  & 10 & 10.0093 \\
-50 mV & 0.794 kV/cm & 1  & 1.0029 \\
 0 mV  &  2.05 kV/cm & 5600 & 5569.4 \\
 0 mV  & 0.9341 kV/cm & 10  & 10.0051 \\
 0 mV  & 0.821 kV/cm & 1 & 1.0099 \\
\end{tabular}
\end{center}
\end{figure}

Figure 3 illustrates the sensitivity the maximum pore number exhibited as the applied electric field was varied.  The applied field for a single pore was aobut .8 kV/cm and just increasing that by a small amount still yields a few pores, less than 50.  However, on the order of doubling the Eapp produces a number of pores approaching 10,000 pores.  Hence, the sensitivity is such that it doesn't take that much change in applied electric field strength to increase the pore number by a huge amount.

\begin{figure}
\begin{center}
\begin{tabular}{c}
\includegraphics[width=3.4in]{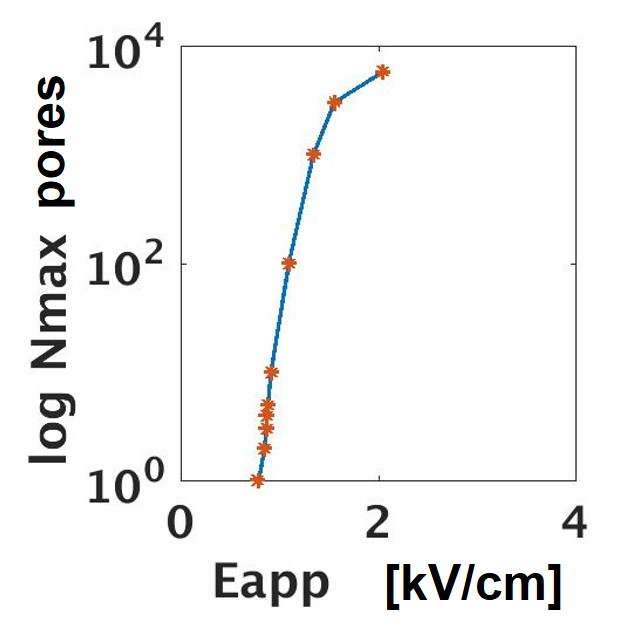} \\ 
  \\
\end{tabular}
\end{center}
\caption{\textbf{Sensitivity to Applied Electric Field.}
Semilog plot of the maximum number of pores as applied electric field is varied.
}
%\end{floatrow}
\end{figure}

\vspace{0.23in}
\Large
\textbf{Results} 
\vspace{0.12in}
\normalsize

Figure 4 shows the transmembrane voltage (Um) for 50 $\mu$s.  This time frame is enough to show the 40 $\mu$s applied pulse and early post-pulse times.  The cathodic and anodic results are shown comparing the "Kennedy pulse" at -50 mV and 0 mV initial resting potential to the single pore case.  The -50 mV and 0 mv initial resting potential results basically overlay each other on the graph for the "Kennedy pulse," but there is a separation for the two curves in the single pore case.  The cathodic early post-pulse region is viewed up close in Figure 5 (left), enlarged to the 38 $\mu$s to 45 $\mu$s time frame encompassing the end of the pulse and the early post-pulse times.  A further enlargement zooms in on the y-axis in Fig 5 (right).  

As can be seen from the figures, the 5700 pore case (dashed curve) immediately reaches 0 mV upon turning off the pulse; hence, full depolarization conditions are met right away.  However, for the single pore case (solid curve), at early
post-pulse times, the transmembrane voltage (Um) relaxes not to 0 mV, but to 
-25 mV.  The single pore does not provide a full depolarization because there is too small a shunt.  

Figure 6 illustrates a similar result for the anodic side.

\begin{figure}
\begin{center}
\begin{tabular}{cc}
\includegraphics[width=2.8in]{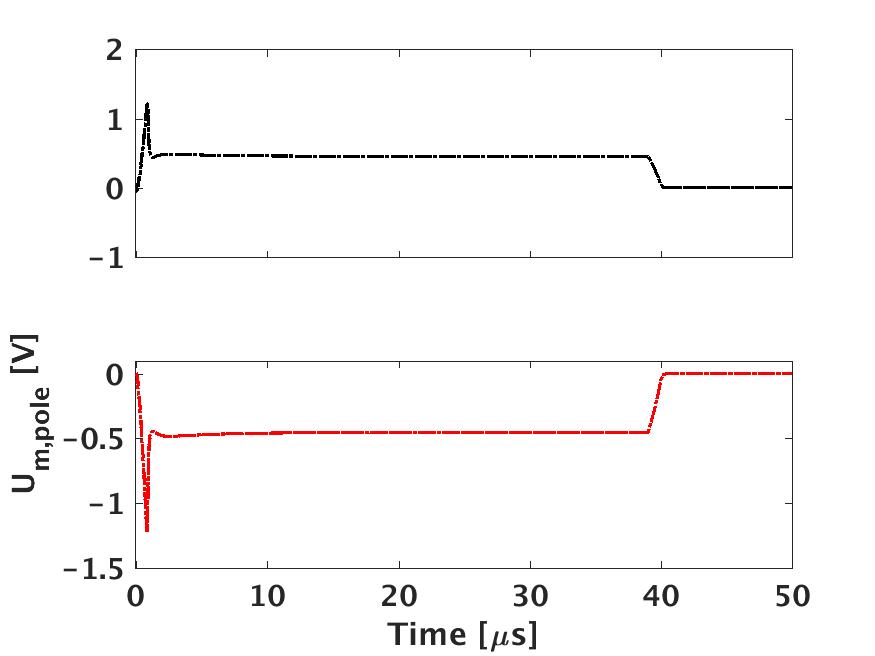} & 
\includegraphics[width=2.8in]{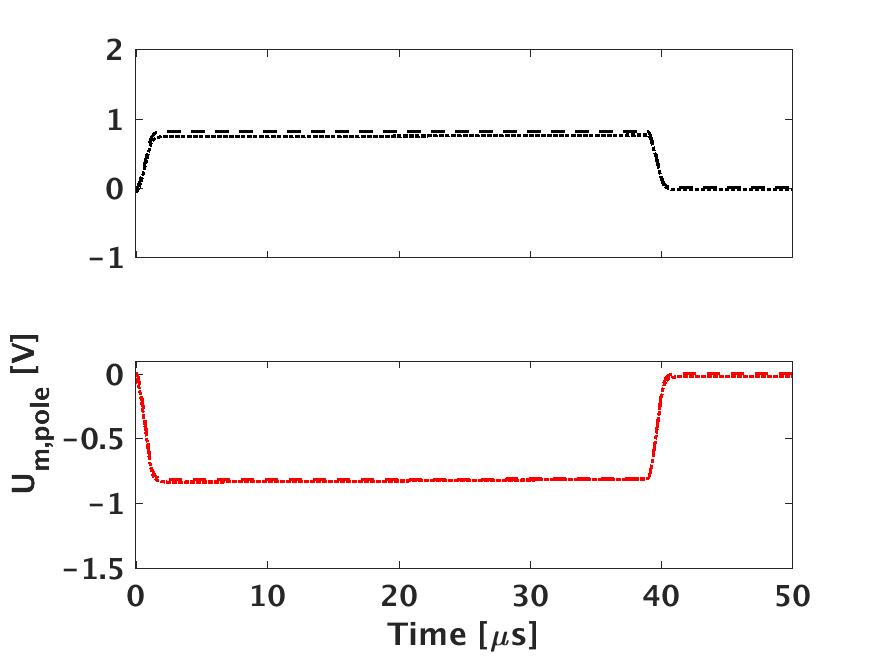} \\
Um 0-50 us & Um 0-50 us  \\
\end{tabular}
\end{center}
\caption{\textbf{Transmembrane Voltage.}
Red is cathodic pole, black is anodic pole.  Dotted is -50mV, dashed is 0mV.  Left upper
and lower are the Kennedy type pulse, 2.05 kV/cm which has 5700 pores for -50mV case,
and 5600 pores for the 0mV case.  The right upper and lower plots are for the
1 pore case and have Eapps of .794 kV/cm for -50mV initial resting potential and .821 kV/cm for 
the 0mV initial resting potential.  
}
\end{figure}

\begin{figure}
\begin{center}
\begin{tabular}{cc}
\includegraphics[width=2.8in]{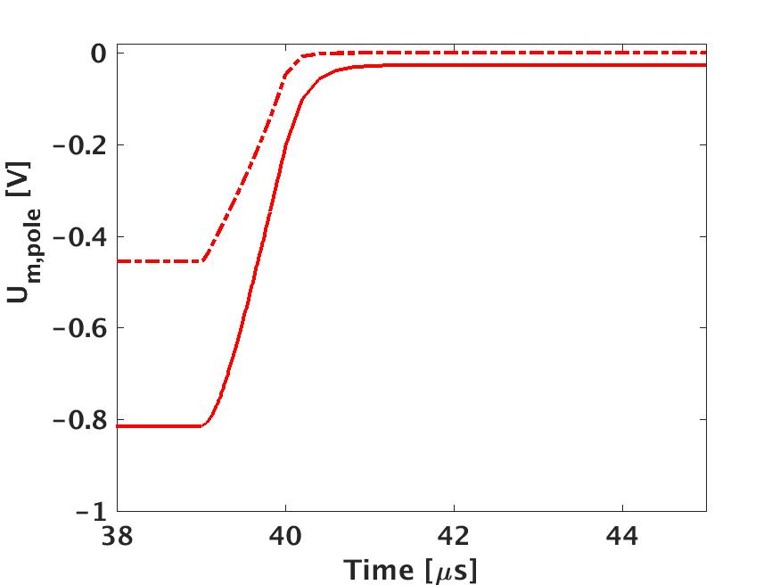} & 
\includegraphics[width=3.4in]{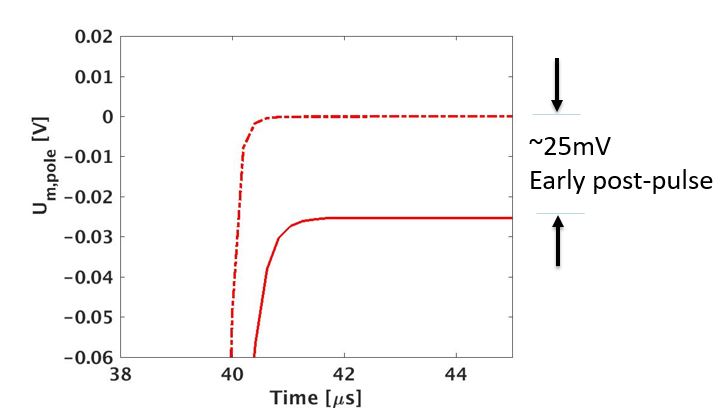} \\
 &   \\
\end{tabular}
\end{center}
\caption{\textbf{Transmembrane Voltage Early Post-Pulse Cathodic.}
Initial $\sim$25mV difference between the 5700 pore case and the 1 pore case.
Dashed is -50mV and 5700 pores.  Solid is -50mV and 1 pore.  Right figure
is a zoom in of the left figure.
}
\end{figure}

\begin{figure}
\begin{center}
\begin{tabular}{cc}
\includegraphics[width=2.8in]{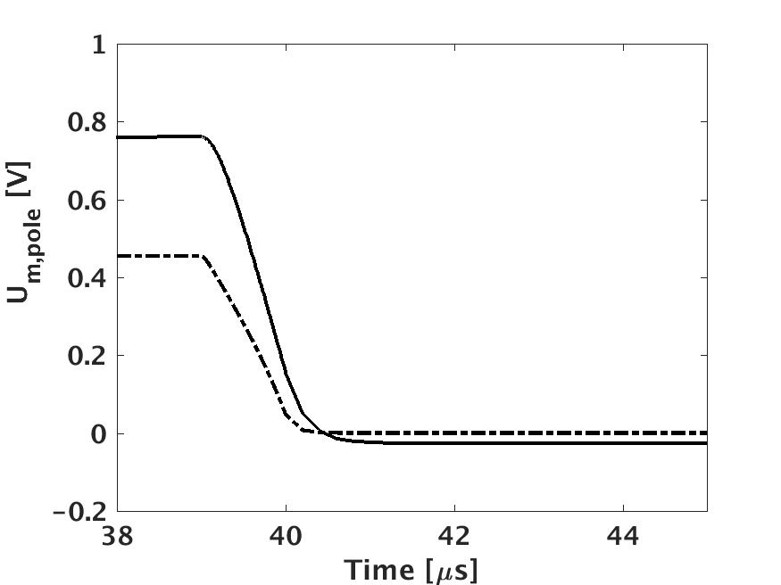} & 
\includegraphics[width=3.4in]{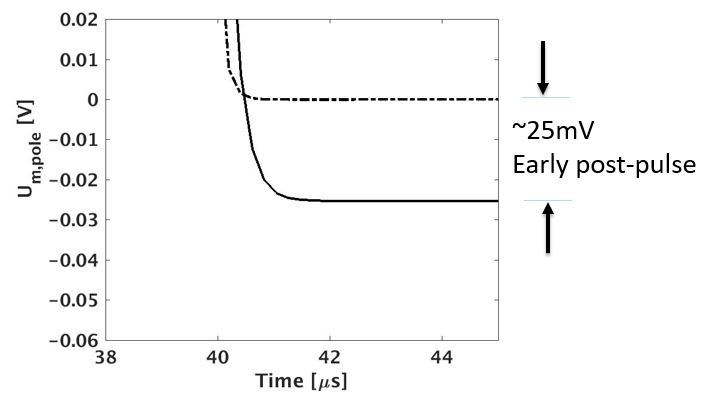} \\
 &   \\
\end{tabular}
\end{center}
\caption{\textbf{Transmembrane Voltage Early Post Pulse Anodic.}
Initial $\sim$25mV difference between the 5700 pore case and the 1 pore case.
Dashed is -50mV and 5700 pores.  Solid is -50mV and 1 pore. Right figure
is a zoom in of the left figure.
}
\end{figure}

The uptake rate of Pr++ was determined by taking $\frac{dn}{dt}$ where n is the number of Pr++ molecules entering the intracellular region of the cell.  Figure 7 illustrates that at the end of the pulse there is a sharp drop in Pr++ uptake.  However, it does not drop to 0, but remains constant.  The 0 mV case is a diffusion only case.  The 5700 and 5600 pores for the -50 mV and 0 mV initial resting potential cases respectively, illustrate no difference between the diffusion only 0 mV case and the -50mV case. This is indicative that in the -50 mV and 5700 pores result, diffusion is dominant during early post-pulse time frames.  However, the figure also shows that for the 1 and 10 pore cases the non-zero constant uptake rate is larger for the -50 mV initial resting potential case than it is for the 0 mV initial resting potential case.  This is an indication that the mechanism of uptake is not only diffusion.

\begin{figure}
\begin{center}
\begin{tabular}{c}
\includegraphics[width=3.8in]{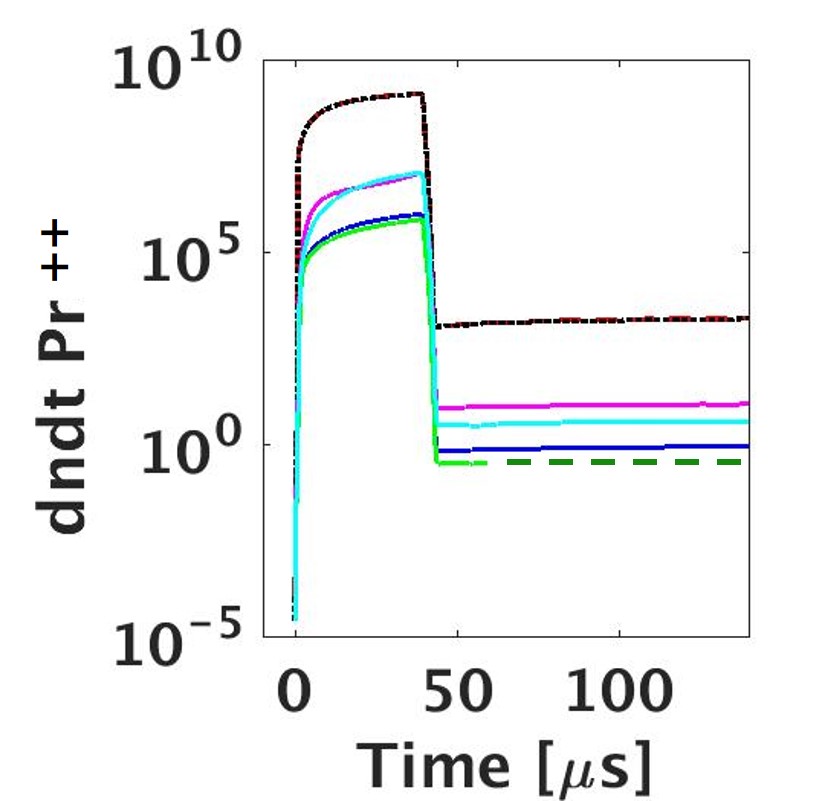} \\
\\
\end{tabular}
\end{center}
\caption{\textbf{Rate of Uptake of Pr++.}
Semilog plots, log10 of dndt is taken. Red, magenta, blue are -50mV for
5700, 10 and 1 pore respectively.  Black, cyan, and green are 0 mV
for 5600, 10, and 1 pore respectively.  Green stops short due to a numerical
error and is under testing.  A line of extrapolation fills in the estimated green line result. 
}
\end{figure}

The pore density distribution is shown in Figure 8.  For the 5700 pores, the anodic and cathodic poles have roughly equal numbers of pores present.  However, for the single pore case, there is clear preference for the anodic pole.  There is a slight difference between the two poles in the 5700 case and the small contribution of the cathodic pole in the single pore case. This may simply be due to the asymmetry brought about by the initial resting potential being negative. This means that cell interior becomes more negative on the cathodic side where the transmembrane voltage adds a negative contribution due to charge separation, and less negative on the anodic side where the transmembrane voltage adds a positive contribution.

\begin{figure}
\begin{center}
\begin{tabular}{cc}
\includegraphics[width=3.4in]{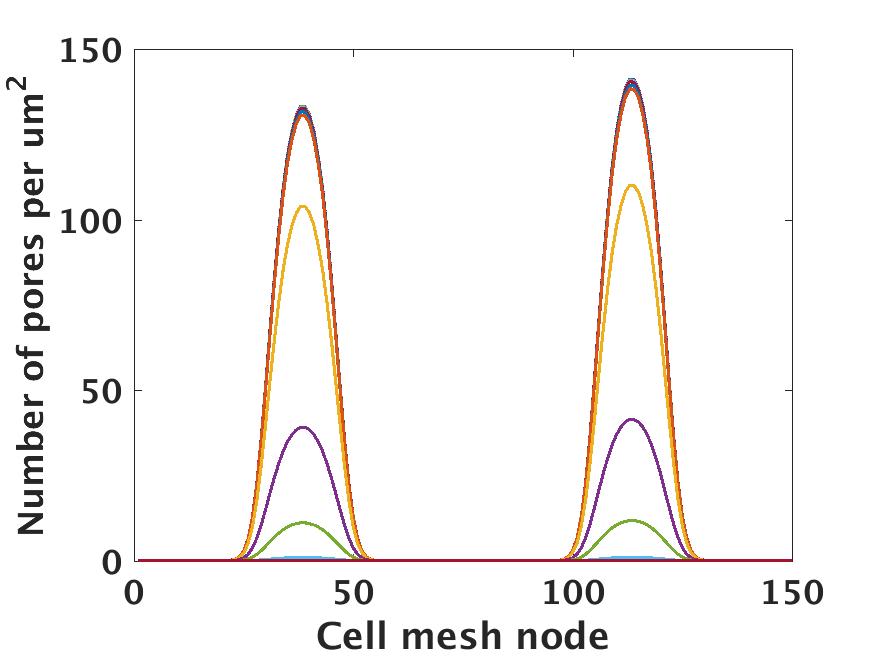} & 
\includegraphics[width=3.4in]{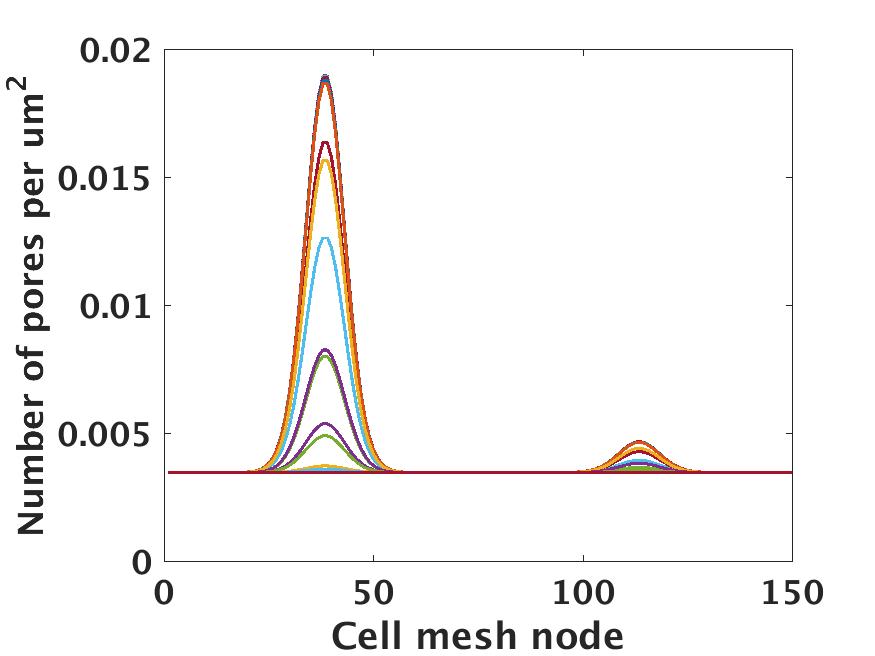} \\ 
 5700 Pores &  1 Pore   \\
\end{tabular}
\end{center}
\caption{\textbf{Pore Density Distributions.}
Pore density distribution plotted over the 150 cell mesh node pairs.  The pore density is
determined as the number of pores per $\mu{m^2}$.  The single pore case appears on the right and the 5700 pore case is on the left. The multiple curves show various time snapshots. The initial resting potential is -50 mV for both.
%determined as the number of pores per $\mum^2$.  The single pore case appears on the right and the 5700 pore case is on the left. The multiple curves show various time snapshots. The initial resting potential is -50 mV for both.
}
\end{figure}

\begin{figure}
\begin{center}
\begin{tabular}{cc}
\includegraphics[width=3.4in]{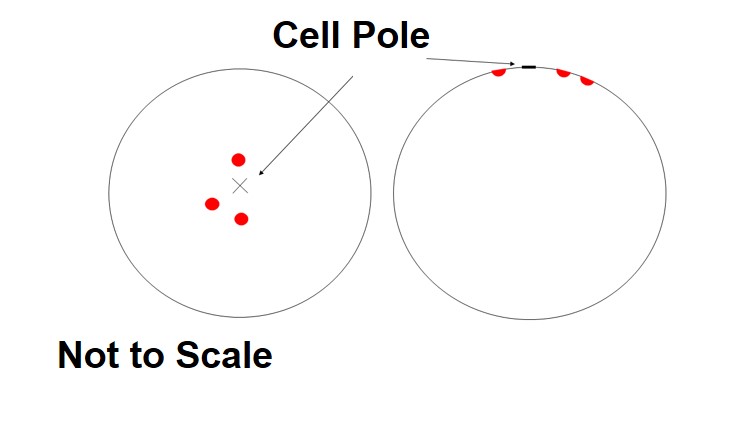} & 
\includegraphics[width=3.4in]{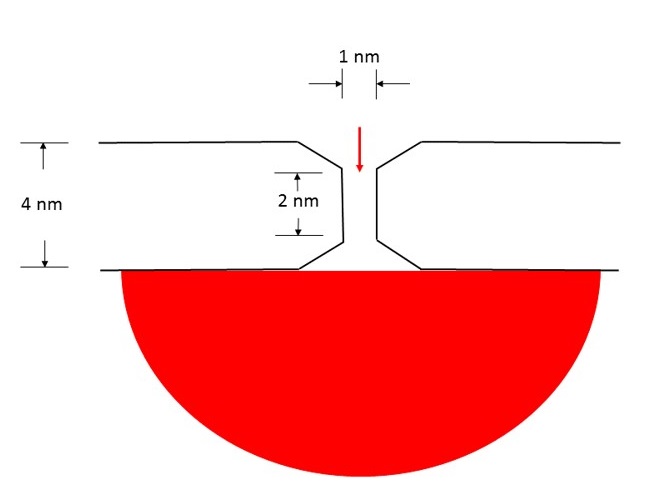} \\ 
 &   \\
\end{tabular}
\end{center}
\caption{\textbf{Schematic of Experimental Detection of a Few Pores.}
Conceptual experimental detection possibility of a few local pores using optical imaging methods.
}
\end{figure}

%--------------------------------------------------------------------------------
\vspace{0.23in}
\Large
\textbf{Discussion}
\vspace{0.12in}
\normalsize

It was shown that single pores do not fully discharge the membrane.  The $\sim$25 mV difference in the early post-pulse time frame illustrate that there is a component of active Propidium transport, electrophoretic drift.  Drift is occurring as the dominant component early post-pulse, but passive diffusion is also taking place.  

The electrodiffusive Pr++ uptake rate drops sharply at the pulse end but does not reach zero.  The constant non-zero post-pulse uptake rate is greater for
cells with -50 mV resting potential than it is for cells with the 0 mV resting potential.  Since the 0 mV case represents diffusion, the non-zero uptake for
this case illustrates the passive transport contribution.  However,
when the uptake rate is constant but greater than this for the -50 mV resting potential, the additional uptake is going beyond passive diffusion with the addition of an active transport component. Between this and the transmembrane voltage results, the additional active transport component is drift.

This is consistent with the findings in S\"{o}zer et. al. 2017 
 \cite{SozerLevineVernier_QuantitativeLimits_SmallMoleculeTransport_Electropermeome_MeasuringModeling-SingleNanosecondPerturbations_SciRep2017}.

The single pore from the pore density distribution curves is appearing and more likely to appear
on the anodic side of the cell.  However, we have seen similar results on the cathodic side as well.  It has been well known that pores can be anywhere and still influence the transmembrane voltage over the entire cell membrane \cite{DeBruinKrassowska_TheoreticalModel_SingleCellEporeI_FieldStrength_RestPotential_BPJ1999}.

It is our contention that an experimentally designed optical imaging method
can find localized regions of a few pores utilizing electric field strength parameters from the model.  In this scenario, we know that the hemispherical regions of interest right on the inner side of the membrane are localized regions of Pr++ transport.  Initially, a certain amount of Pr++ 
will flow through the pore and bind.  A voltage-sensitive dye will determine the locale of these high impact
 Pr++ regions, and with an imaging method to detect them, a single pore or a few pores can simply be counted visually.  The model can provide the applied electric field strength needed to generate the single or few pores.  The model can also provide a start time for when to look, commencing early post-pulse.  But, since the pores stay around for awhile and the uptake is constant, there is a range of times where the pores may be visually counted.

\vspace{0.23in}
\Large
\textbf{Acknowledgments}
\vspace{0.12in}
\normalsize

This work was supported by AFOSR MURI grant FA9550-15-1-0517 on Nanoelectropulse-Induced Electromechanical Signaling and Control of Biological Systems, administered through Old Dominion University. 

\vspace{0.12in}
We thank K. G. Weaver for computer support.

\vspace{0.32in}
\textbf{References}
\vspace*{-0.52in}
\small
\def\refname{}

\bibliographystyle{unsrt}
\normalsize

\normalsize

\end{document}